\documentclass[preprint,showpacs,preprintnumbers,amsmath,amssymb]{revtex4}
\usepackage{graphicx}
\usepackage{dcolumn}
\usepackage{bm}
\begin{document}
\title {Lifetime of a quasiparticle in an electron liquid. II}
\author {Zhixin Qian}
\affiliation{Department of Physics,
Peking University, Beijing 100871, China}
\date{\today}
\begin{abstract}
Inelastic lifetime of an electron quasiparticle in an
electron liquid due to
electron-electron interaction evaluated in previous work
is calculated in an alternative way. Both the contributions
of the ``direct" and ``exchange" processes are included. 
The results turn out to be exactly the same as those
obtained previously, and hence confirm the latter and
consequently fully resolve the theoretical
discrepancies existing in the literature. Derivation 
in the two-dimensional
case is presented in detail due to its intricacies. 
The effects of local field and finite well width on
the effective electron interaction in the two-dimensional case are also
investigated in a quantitive comparison of the electron
relaxation rate between theory and experiment. These
effects are shown to make rather small contribution
to the quasiparticle lifetime.
\end{abstract}
\maketitle

\section{Introduction}

Low-energy electron excitations in solids can be successfully
described in terms of quasiparticles in the Landau theory
of Fermi liquids.\cite{pines,GV} An excited quasiparticle with definite
momentum is not stable due to scattering by
phonons, disorders, and other electrons. Hence a quasiparticle
has finite lifetime. Among this, the intrinsic
inelastic scattering lifetime $\tau_e$,
i.e., the lifetime that arises purely from the
electron-electron scattering processes, is a central quantity
in the Laudau theory of the electron
liquid. It plays a key role
in our understanding of a broad variety of phenomena in
solids such as
electron dephasing,\cite{dephase} tunneling,\cite{tunneling,murphy}
and localization,\cite{local} etc. It might also have effect
on electron transport.\cite{transport}

In fact, the electron tunneling techniques in semiconductor quantum
wells \cite{tunneling,murphy} have enabled 
experimentalists to directly determine $\tau_e$
in two-dimensional (2D) electron liquids. 
For weakly coupled wells, the lifetime 
principally arises from electron-electron scattering processes.
On the other side, huge progress has also been made, by the
use of the techniques of ultrafast laser, in measuring the lifetime
of photonexcited electrons in metals such as copper.\cite{ogawa}
These advances have made it possible to carry out quantitive comparisons 
between theories and experiments. The theory of the 
inelastic lifetime in three dimensions (3D) 
is rather well established within random
phase approximation.\cite{quinn,ritchie} 
It was later extended to include the exchange 
contribution (see Ref. \cite{QV} for a
detailed discussion).\cite{kleinman,penn}
Several theoretical investigations had also been carried out in
2D, but with quantitive
disagreement.\cite{giuliani,fukuyama,chaplik,hodges,jungwirth,zheng,reizer} 
In an earlier paper,\cite{QV} hereafter referred as I, we have
managed to clarify the origin of the 
disagreement that exists among these previous
investigations. The results in I are summarized as follows.
The inverse lifetime of a quasiparticle with low
energy $\xi_p$ (relative to the chemical potential $\mu$) at
temperature $T$ in a 3D electron liquid is
\begin{eqnarray}  \label{tau3Dfina}
\frac{1}{\tau_e}
= && \frac{m^3 e^4}{\pi p k_s^3} \frac{\pi^2 k_B^2T^2 +\xi_p^2}
{1+e^{- \beta \xi_p}} \biggl [ \frac{\lambda}{\lambda^2 +1}
+ \tan^{-1} \lambda  \nonumber \\
&& - \frac{1}{\sqrt{\lambda^2 +2 }}
\biggl \{ \frac{\pi}{2} - \tan^{-1}
\biggl ( \frac{1}{\lambda}
\sqrt{\frac{1}{\lambda^2 +2 }} \biggr ) \biggr \} \biggr ] ,
\end{eqnarray}
where $\lambda =2k_F/k_s$, and 
$k_s = \sqrt{\frac{4k_F}{\pi a_0}}$ is the
3D Thomas-Fermi screening wavevector. $k_B$, $k_F$, and
$a_0$ are the Boltzmann constant, the Fermi wavevector, and the Bohr
radius, respectively. For a 2D electron liquid, we found
\begin{eqnarray}  \label{taueeT=0}
\frac{1}{\tau_e} =  -\frac{m^2 \xi_p^2}{16 \pi^3 E_F}
[ && 3 W^2(0) +  2  W^2(2k_F)   \nonumber \\
&& - 2 W(0) W(2k_F) ] \ln \frac{\xi_p}{2 E_F}~
\end{eqnarray}
for $k_BT \ll \xi_p$,
and 
\begin{eqnarray} \label{taueeT>0}
\frac{1}{\tau_e}
= -\frac{\left(m k_B T\right)^2}{32 \pi E_F}
[&& 3 W^2(0) + 2  W^2(2k_F) \nonumber \\ 
&& - 2 W(0) W(2k_F) ]
\ln \frac{k_B T}{2E_F}~
\end{eqnarray}
for $\xi_p \ll k_BT$. Here $E_F=\hbar^2 k_F^2/2m$, and
$W(q)$ is the effective interaction between quasiparticles.

The calculation of $\tau_e$ is a quite nontrivial task in many-body
theory, which helps explain the disagreement among various
previous theoretical results in 2D. Evidently, correctness of the
results in Eq. (\ref{tau3Dfina}), and Eqs. (\ref{taueeT=0}) and
(\ref{taueeT>0}) is crucial in any meaningful comparisons with experiments.
In fact, big discrepancies remain between experiments
and theories, and call for 
explanations.\cite{tunneling,murphy,ogawa,QV,giuliani,fukuyama,chaplik,hodges,jungwirth,zheng,reizer}
In this paper, we calculate 
$1/\tau_e$ in an alternative way, and  
confirm the above results. 
Various theoretical discrepancies in the literature, as mentioned
above, are fully resolved.

The present calculation appears 
very different to the previous calculation
in I.
In I, we calculated $1/\tau_e$ by expressing it as the frequency
convolution of the imaginary part of the density-density response
function.
The present calculation is
technically straightforward, and somehow in a textbook
fashion. But it is by no means much simpler than that in I.
In fact, the present approach seems rather clumsy
for the case of $\xi_p \ll k_BT$.
Hence we shall restrict ourselves to the case of zero temperature.

After giving the general formulas for $1/\tau_e$ in the next section,
we present our calculation for the 3D and 2D cases separately
in Sec. III and IV. In Sec. V, we shall discuss the contribution
to the inverse quasiparticle lifetime in 2D arising from the effects of
local field and finite well width on the effective electron
interaction, and then briefly
summarize the paper.
 
\section{General Formulas}

We start by rewriting Eqs. (4) and (5) in I as follows:
\begin{eqnarray}
\frac{1}{\tau_\sigma^{(D)}} = && 2 \pi \sum_{{\bf k}, {\bf q}}
\sum_{\sigma'} W_{\sigma \sigma'}^2({\bf k} - {\bf p}) 
\bar n_{{\bf k} \sigma}
n_{{\bf k} + {\bf q} -{\bf p} \sigma'} \bar n_{{\bf q} \sigma'} 
\nonumber \\
&& \delta( \xi_{{\bf p} } + \xi_{{\bf k} + {\bf q} - {\bf p}
\sigma'}  - \xi_{{\bf k} \sigma}- \xi_{{\bf q} \sigma'}),
\end{eqnarray}
and
\begin{eqnarray}
&&\frac{1}{\tau_{\sigma}^{(ex)}} = - 2 \pi \sum_{{\bf k}, {\bf q}}
W_{\sigma \sigma}({\bf p} - {\bf q})
W_{\sigma \sigma}({\bf k} - {\bf p}) \nonumber \\ 
&& \bar n_{{\bf k} \sigma}
n_{{\bf k} + {\bf q} -{\bf p} \sigma'} \bar n_{{\bf q} \sigma'} 
\delta( \xi_{{\bf p}} + \xi_{{\bf k} + {\bf q} - {\bf p}
\sigma} - \xi_{{\bf k} \sigma}- \xi_{{\bf q} \sigma}), 
\end{eqnarray}
where $W_{\sigma \sigma'}({\bf q})$ is the effective interaction
between quasiparticles of spin $\sigma$ and spin $\sigma'$. We have
set $\hbar=1$. We consider only the paramagnetic 
electron liquid, and hence the index $\sigma$
of the $1/\tau_\sigma^{(D), (ex)}$ is
unnecessary and will be
dropped hereafter.
The sum of $1/\tau^{(D)}$ and $1/\tau^{(ex)}$ yields the inverse
inelastic lifetime $1/\tau_e$,
\begin{eqnarray}
\frac{1}{\tau_e} = 2 \pi \sum_{{\bf k}, {\bf q}}
\sum_{\sigma'} && (1- \frac{1}{2} \delta_{\sigma \sigma'})
[W_{\sigma \sigma'}({\bf k} - {\bf p})   \nonumber \\ 
&& -\delta_{\sigma \sigma'}W_{\sigma \sigma'}
({\bf p} - {\bf q})]^2  
\bar n_{{\bf k} \sigma}
n_{{\bf k} + {\bf q} -{\bf p} \sigma'} \bar n_{{\bf q} \sigma'} 
\nonumber \\
&& \delta( \xi_{\bf p} + \xi_{{\bf k} + {\bf q} - {\bf p}
\sigma'} - \xi_{{\bf k} \sigma} - \xi_{{\bf q} \sigma'}).
\end{eqnarray}

We are only interested in the case of low excited energy 
($ \xi_p \ll E_F$). In this case the contribution to
the summations over momenta in the 
above expression arises only from the region of
$\xi_{{\bf k} + {\bf q} - {\bf p} \sigma}$, $\xi_{{\bf k} \sigma}$,
$\xi_{{\bf q} \sigma} \ll E_F$. Therefore
\begin{eqnarray}  \label{geneform}
\frac{1}{\tau_e} =  2 \pi && \sum_{{\bf k}, {\bf q}}
\sum_{\sigma'} 
I^{\sigma \sigma'}(\mu_{\bf k}, \mu_{\bf q})
\bar n_{{\bf k} \sigma}
n_{{\bf k} + {\bf q} -{\bf p} \sigma'} \bar n_{{\bf q} \sigma'}
\nonumber \\
&& \delta( \xi_{{\bf p} \sigma} + \xi_{{\bf k} + {\bf q} - {\bf p}
\sigma'} - \xi_{{\bf k} \sigma} - \xi_{{\bf q} \sigma'}),
\end{eqnarray}
where $\mu_{\bf k} = \hat p \cdot \hat k$ and
$\mu_{\bf q} = \hat p \cdot \hat q$ respectively, and
the hats mean unit vectors.
We have defined formally
\begin{eqnarray}
I^{\sigma \sigma'}(x, y) = && (1 - \frac{1}{2}\delta_{\sigma \sigma'})
[W_{\sigma \sigma'}(\sqrt{2}k_F \sqrt{1 -x}) \nonumber \\
&& -\delta_{\sigma \sigma'}
W_{\sigma \sigma'}(\sqrt{2}k_F \sqrt{1 -y}) ]^2 .  
\end{eqnarray}

Below we present our calculation for the 3D and 2D cases
separately 
in Sec. III and Sec. IV.
We set the volume of the 3D system 
and the area of the 2D system, respectively, to be unit in this paper.

\section{The inverse lifetime in 3D}

The integrations over the azimuthal angles of ${\bf k}$ and ${\bf q}$ 
with respect to ${\bf p}$ in Eq. (\ref{geneform})
can be straightforwardly carried out. 
After that, it becomes
\begin{eqnarray}  \label{3DEq1}
\frac{1}{\tau_e} && =  2 \pi \biggl ( \frac{1}{(2 \pi)^3} \biggr )^2
\frac{\sqrt{2} \pi}{E_p} \sum_{\sigma'}
\int_{k_F}^\infty dk k^2 \int_{k_F}^\infty dq q^2 \nonumber \\
&& \theta(p^2+k_F^2 -k^2-q^2) 
\int_{-1}^1 d \mu_{\bf k} \int_{-1}^1 d \mu_{\bf q}  \nonumber \\
&& I^{\sigma \sigma'}(\mu_{\bf k}, \mu_{\bf q})  
\frac{1}{\sqrt{(1 - \mu_{\bf k}) (1- \mu_{\bf q})}}
\frac{\theta(\mu_{\bf k} 
+ \mu_{\bf q})}{\sqrt{\mu_{\bf k} + \mu_{\bf q}}},
\end{eqnarray}
where $E_p=p^2/2m$.
We have defined $\theta(x)=1$ for $ x>0$, $\theta(x)=0$ for $ x \le 0$.
Notice that Eq. (\ref{3DEq1}) is essentially 
the same as Eq. (4) in Ref. \cite{penn} 
except typos of $(\frac{\Omega}{2 \pi^3})^3$ and a missing factor 
of $1-\frac{1}{2} \delta_{\sigma \sigma'}$ in Eq. (3b) in Ref. \cite{penn}. 
The integrations over $k$ and $q$ in Eq. (\ref{3DEq1}) can be
carried out analytically
to the accuracy of the leading order of $O(\xi_p^2)$, and yield
\begin{eqnarray}
&& \frac{1}{\tau_e} = A \frac{\xi_p^2}{E_F E_p} 
\sum_{\sigma'}
\int_{-1}^1 d \mu_{\bf k} \int_{-1}^1 d \mu_{\bf q}  \nonumber \\
&& I^{\sigma \sigma'}(\mu_{\bf k}, \mu_{\bf q})  
\frac{1}{\sqrt{(1 - \mu_{\bf k}) (1- \mu_{\bf q})}}
\frac{\theta(\mu_{\bf k} 
+ \mu_{\bf q})}{\sqrt{\mu_{\bf k} + \mu_{\bf q}}} ,
\end{eqnarray}
where
\begin{eqnarray}
A= \frac{e^2}{a_0} \biggl (\frac{1}{2^{5/2} \pi^2} \biggr )
\biggl (\frac{k_F^2}{4 \pi e^2} \biggr )^2 .
\end{eqnarray}
Notice that there seems an error of a factor $1/2$ 
in Eq. (6b) in Ref. \cite{penn}.
However this does not effect the results in
Table II in Ref. \cite{penn} and the subsequent conculsions, since
the coefficient $A$ cancels in the 
ratio $p^{\sigma \sigma}/p^{\sigma \bar \sigma}$ in the table.

The contributions from the ``direct" and ``exchange" processes are,
separately, given as,
\begin{eqnarray}    \label{3DDgen}
\frac{1}{\tau^{(D)}} &&= A \frac{\xi_p^2}{E_F E_p} 
\int_{-1}^1 d \mu_{\bf k} \int_{-\mu_{\bf k}}^1 d \mu_{\bf q} \nonumber \\
&& [W_{\sigma \sigma}^2(\sqrt{2}k_F \sqrt{1- \mu_{\bf k}}) +
W_{\sigma {\bar \sigma}}^2(\sqrt{2}k_F \sqrt{1- \mu_{\bf k}}) ]  \nonumber \\ 
&& \frac{1}{\sqrt{(1 - \mu_{\bf k}) (1- \mu_{\bf q})}}
\frac{1}{\sqrt{\mu_{\bf k} + \mu_{\bf q}}} ,
\end{eqnarray} 
where ${\bar \sigma}=- \sigma$, and
\begin{eqnarray}   \label{3Dexgen}
\frac{1}{\tau^{(ex)}} =&& -A \frac{\xi_p^2}{E_F E_p} 
\int_{-1}^1 d \mu_{\bf k} \int_{-\mu_{\bf k}}^1 d \mu_{\bf q} \nonumber \\
&& W_{\sigma \sigma}(\sqrt{2}k_F \sqrt{1- \mu_{\bf k}})  
W_{\sigma \sigma}(\sqrt{2}k_F \sqrt{1- \mu_{\bf q}}) \nonumber \\ 
&& \frac{1}{\sqrt{(1 - \mu_{\bf k}) (1- \mu_{\bf q})}}
\frac{1}{\sqrt{\mu_{\bf k} + \mu_{\bf q}}} .
\end{eqnarray}

The spin dependence of the effective interaction is not
crucial in determining the total inelastic lifetime.\cite{penn}
We shall ignore this dependence and follow the usual practice
of characterizing the screening effects by the screening wavevector 
$k_s$:
\begin{eqnarray}  \label{TF}
W(q) = \frac{4 \pi e^2}{q^2 +k_s^2} .
\end{eqnarray}
The integrations over $\mu_{\bf q}$ and $\mu_{\bf k}$ can be 
analytically carried out.
After that, one finally has
\begin{eqnarray}  \label{tau3D}
\frac{1}{\tau^{(D)}}
= \frac{m^3 e^4k_F}{\pi p^2 k_s^3} \xi_p^2
\biggl [ \frac{\lambda}{\lambda^2 +1}
+ \tan^{-1} \lambda \biggr ] ,
\end{eqnarray}
and
\begin{eqnarray}  \label{tau3ex}
\frac{1}{\tau^{(ex)}}
&&=-\frac{m^3 e^4k_F}{ \pi p^2 k_s^3} \xi_p^2
\frac{1}{\sqrt{\lambda^2 +2 }}  \nonumber  \\
&&\biggl [ \frac{\pi}{2} - \tan^{-1}
\biggl ( \frac{1}{\lambda}
\sqrt{\frac{1}{\lambda^2 +2 }} \biggr ) \biggr ] .
\end{eqnarray}
Equations (\ref{tau3D}) and (\ref{tau3ex}) 
are, to the leading order of $O(\xi_p^2)$, 
exactly the same as the results for $1/\tau^{(D)}$ and $1/\tau^{(ex)}$
obtained in I for the 3D case.

\section{The inverse lifetime in 2D}

The derivation in the 2D case is relatively intricate. 
We shall present it in details.
We first rewrite Eq. (\ref{geneform}) as
\begin{eqnarray}
&& \frac{1}{\tau_e} = 2 \pi m \biggl 
(\frac{1}{(2 \pi)^2} \biggr )^2
\sum_{\sigma'} \int_{k_F}^\infty dq q 
\int_{k_F}^\infty dk k   \nonumber \\
&&\theta(p^2 + k_F^2 - k^2 - q^2) 
\int_{- \pi}^\pi d \phi_{\bf k} \int_{- \pi}^\pi d \phi_{\bf q}
I^{\sigma \sigma'}(\mu_{\bf k}, \mu_{\bf q}) \nonumber \\
&& \delta(p^2 + kq \cos(\phi_{\bf k} - \phi_{\bf q})
- pk \cos \phi_{\bf k} - pq \cos \phi_{\bf q} ) , 
\end{eqnarray}
where $\mu_{\bf k}= cos \phi_{\bf k}$, $\mu_{\bf q}= cos \phi_{\bf q}$,
and $\phi_{\bf k}$ and $\phi_{\bf q}$ are the angles of
${\bf k}$ and ${\bf q}$ relative to ${\bf p}$, respectively.
After some algebraic manipulations, the
preceding expression can be written as
\begin{eqnarray}   \label{2DEq2}
\frac{1}{\tau_e} &&= 8 \pi m \biggl ( \frac{1}{(2 \pi)^2} \biggr )^2
\sum_{\sigma'} 
\int_{k_F}^p dq q \int_{k_F}^{\sqrt{p^2 + k_F^2- q^2}} dk k 
\nonumber \\
&& \int_{- 1}^1 d \mu_{\bf k} \int_{- 1}^1 d \mu_{\bf q}
I^{\sigma \sigma'}(\mu_{\bf k}, \mu_{\bf q}) \nonumber \\
&& \delta [(p - k\mu_{\bf k})^2 (p - q\mu_{\bf q})^2
-k^2q^2(1- \mu_{\bf k}^2)(1- \mu_{\bf q}^2) ] .  \nonumber \\
\end{eqnarray}
We now define the following variables:
\begin{eqnarray}
\lambda=\frac{p}{\sqrt z}, ~~ \lambda'=\frac{p}{\sqrt{z'}} ,
\end{eqnarray}
and function:
\begin{eqnarray}  \label{f(xy)}
f(x, y)= (\lambda - x)^2(\lambda' - y)^2 - 1 +x^2 +y^2 -x^2y^2,
\end{eqnarray}
\begin{figure}
\unitlength1cm
\begin{picture}(5.0,6.0)
\put(-5.0,-4.0){\makebox(7.0,8.0){
\includegraphics{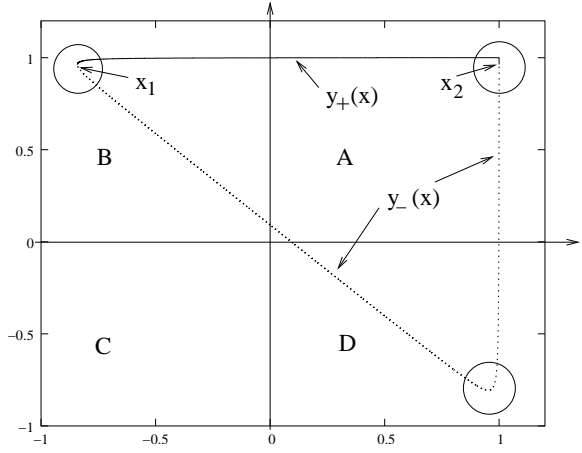}
}}
\end{picture}
\caption{The contour of $f(x, y) =0$, defined in Eq. (\ref{f(xy)}), 
and its two
parts: $y_+(x)$ (solid line) and $y_-(x)$ (dotted line), defined
in Eq. (\ref{ypm}).}
\label{figur1}
\end{figure}
and write Eq. (\ref{2DEq2}) in the following form:
\begin{eqnarray}
\frac{1}{\tau_e} = && \frac{2 \pi m}{k_F^2}
\biggl ( \frac{1}{(2 \pi)^2} \biggr )^2 \sum_{\sigma'}
\int_{k_F^2}^{p^2} dz \int_{k_F^2}^{p^2+k_F^2-z} dz'  \nonumber \\
&& \int_{-1}^{1} dx \int_{-1}^{1} dy
\delta[f(x, y)]
I^{\sigma \sigma'}(x, y) .
\end{eqnarray}
The contour of $f(x, y)=0$ is illustrated in Fig. 1.

We further define
\begin{eqnarray}  \label{G-J}
G_{\sigma \sigma'}^{A, B, D} = \int_{k_F^2}^{p^2} dz 
\int_{k_F^2}^{p^2+k_F^2-z} dz'
J_{\sigma \sigma'}^{A, B, D} ,
\end{eqnarray}
where
\begin{eqnarray}  \label{JA}
J_{\sigma \sigma'}^A = \int_{0}^{1} dx \int_{0}^{1} dy
\delta[f(x, y)]
I^{\sigma \sigma'}(x, y) ,
\end{eqnarray}
\begin{eqnarray}  \label{JB}
J_{\sigma \sigma'}^B = \int_{-1}^{0} dx \int_{0}^{1} dy
\delta[f(x, y)] 
I^{\sigma \sigma'}(x, y) ,
\end{eqnarray}
and
\begin{eqnarray}  \label{JD}
J_{\sigma \sigma'}^D = \int_{0}^{1} dx \int_{-1}^{0} dy
\delta[f(x, y)]
I^{\sigma \sigma'}(x, y) ,
\end{eqnarray}
and express $1/\tau_e$ as follows:
\begin{eqnarray}  \label{tau-G}
\frac{1}{\tau_e} = \frac{2 \pi m}{k_F^2}
\biggl ( \frac{1}{(2 \pi)^2} \biggr )^2 \sum_{\sigma'}
[G_{\sigma \sigma'}^A + G_{\sigma \sigma'}^B + 
G_{\sigma \sigma'}^D ] .
\end{eqnarray}
Evidently, $A$, $B$ and $D$ denote the contributions
arising from the first, the second and the fourth quadrants
respectively, as shown in Fig. 1. Notice that $f(x, y) \neq 0$
in the third quadrant. In fact,
the leading order contributions
only arise from the circled regions in Fig. 1.
By interchanging the integral variables $x$ and $y$
in Eq. (\ref{JD}), it is easy to see that 
the evaluation of $G_{\sigma \sigma'}^D$ is
totally analogous to that of $G_{\sigma \sigma'}^B$. Hence we only
present the latter.
The evaluation of $G_{\sigma \sigma'}^B$ turns out to be relatively simpler
than that of $G_{\sigma \sigma'}^A$,
and it is in the meanwhile instructive for the latter.
We hence start with the former below. 

We first evaluate the integrations over the variables
$x$ and $y$ in Eq. (\ref{JB}). To this end,
we rewrite Eq. (\ref{JB}) as
\begin{eqnarray} \label{JB5}
J_{\sigma \sigma'}^B=&&\int_{-1}^{0} dx \int_{0}^{1} dy
\frac{1}{(\lambda -x )^2 +1 -x^2}  \nonumber \\
&& \delta([y-y_+(x)][y-y_-(x)])
I^{\sigma \sigma'}(x, y) ,
\end{eqnarray}
where
\begin{eqnarray} \label{ypm}
&& y_{\pm}(x)  =  \frac{1}{(\lambda- x)^2
+1 - x^2}
[\lambda' (\lambda - x)^2   \nonumber \\
&& \pm \sqrt{1-x^2}
\sqrt{(1- \lambda'^2)(\lambda - x)^2 +1- x^2}] .
\end{eqnarray}
The functions $y_+(x)$ and $y_-(x)$ are two components of the
contour of $f(x, y) =0$, and they are illustrated in Fig. 1.
The
integration over the variable $y$ in Eq. (\ref{JB5}) is
now straightforward, which yields
\begin{eqnarray}
J_{\sigma \sigma'}^B = && \frac{1}{2}\int_{-1}^0 dx
\frac{\theta(x - x_1)\theta(x_2 - x)}
{\sqrt{1-x^2}\sqrt{(1-\lambda'^2)(\lambda - x)^2
+1 - x^2}}    \nonumber \\ 
&& [ I^{\sigma \sigma'}(x, y_+(x))
+ I^{\sigma \sigma'}(x, y_-(x)) ] ,
\end{eqnarray}
or, more explicitly,
\begin{eqnarray}
J_{\sigma \sigma'}^B = && \frac{1}{2 \lambda'}\int_{x_1}^0 dx
\frac{1}{\sqrt{(1-x^2)(x_2 - x)(x -x_1)}}  \nonumber \\ 
&& [ I^{\sigma \sigma'}(x, y_+(x))
+ I^{\sigma \sigma'}(x, y_-(x)) ] .
\end{eqnarray}
Here we have defined
\begin{eqnarray}
x_{1,2} = - \frac{\lambda(1-\lambda'^2) \pm \sqrt{\lambda^2
+\lambda'^2 - \lambda^2 \lambda'^2}}{\lambda'^2} ,
\end{eqnarray}
which are also shown in Fig. 1.
By using the fact that $x_1=4(\lambda' -1 )-1$ and
$y_+(x_1) = y_-(x_1) =1$ for $\lambda' \to 1$,
we have, to the leading order,
\begin{eqnarray} \label{JB2}
J_{\sigma \sigma'}^B = - \frac{1}{2} I^{\sigma \sigma'}(-1, 1)
\ln [\lambda' -1] .
\end{eqnarray}
Substituting
the preceding result into 
Eq. (\ref{G-J}) and performing the integrations over $z$ and $z'$,
one obtains
\begin{eqnarray}  \label{GB}
G_{\sigma \sigma'}^B = - \frac{1}{4} I^{\sigma \sigma'}(-1, 1)
(p^2 - k_F^2)^2 \ln[(p^2 - k_F^2)/2k_F^2] .
\end{eqnarray}

As pointed out previously, the evaluation of $G_{\sigma \sigma'}^D$
is similar to that of $G_{\sigma \sigma'}^B$. Here we only
quote the final result,
\begin{eqnarray} \label{GD}
G_{\sigma \sigma'}^D = - \frac{1}{4} I^{\sigma \sigma'}(1, -1)
(p^2 - k_F^2)^2 \ln[(p^2 - k_F^2)/2k_F^2] .
\end{eqnarray}


Next we calculate $G_{\sigma \sigma'}^A$.
First of all, from the experience in deriving $J_{\sigma \sigma'}^B$
in Eq. (\ref{JB2}),
it is not difficult to see that the leading order contribution
to $J_{\sigma \sigma'}^A$ arises from the region of $x \to 1$, $y \to 1$.
Therefore, we may directly rewrite Eq. (\ref{JA}) as
\begin{eqnarray}   \label{JA0}
J_{\sigma \sigma'}^A = I^{\sigma \sigma'}(1, 1)
\int_0^1 dx \int_0^1 dy \delta [f(x, y)] .
\end{eqnarray}
However, the following calculation is a little more
delicate.  Due to the fact
that $y_-(x)$ becomes ill-defined (actually becomes $x=1$)
in the first quadrant as $\lambda$, $\lambda' \to 1$,
a straightforward calculation like the preceding one for
$J_{\sigma \sigma'}^B$ does not work.
To circumvent this difficulty, we 
make the following variable transform,
\begin{eqnarray}
x = \frac{1}{\sqrt{2}}(x' - y') , ~~~ y = \frac{1}{\sqrt{2}}
\frac{\lambda'}{\lambda}(x' + y') .
\end{eqnarray}
and rewrite Eq. (\ref{JA0}) as
\begin{eqnarray}  \label{JA2}
J_{\sigma \sigma'}^A = &&  I^{\sigma \sigma'}(1, 1)
\int_0^{\frac{1}{\sqrt{2}}(1 + \lambda/\lambda')}
dx' \int d y'   \nonumber \\
&& \delta (a(x') [ y'-y_1(x')][y' - y_2(x')]) .
\end{eqnarray}
The Jacobian $\lambda'/\lambda$ for the above
integration
variable transform can be
set to be one in the limit of
$\lambda$, $\lambda' \to 1$.
The limits of the integration over $y'$
are left unspecified because they are not really relevant
simply due to the $\delta$-function
in the integrand, while 
the integration region of $x'$ and $y'$ corresponds to the square
of $0 \le x', y' \le 1$.   
In Eq. (\ref{JA2}), we have defined
\begin{eqnarray}
y_{1,2}(x)= \frac{-b(x) \pm \sqrt{b^2(x) - 4a(x) c(x)}}{2 a(x)} ,
\end{eqnarray}
where
\begin{eqnarray}
a(x) = \frac{1}{2} [2 \sqrt{2}x\lambda'^2/\lambda - 2 \lambda'^2
+\lambda'^2/\lambda^2 + 1 ] ,
\end{eqnarray}
\begin{eqnarray}
b(x) = - (1 - \lambda'^2/\lambda^2)x ,
\end{eqnarray}
and
\begin{eqnarray}
c(x) && =  \lambda^2 \lambda'^2 - 1 
- 2 \sqrt{2} \lambda \lambda'^2 x   \nonumber \\
&& + \frac{1}{2}(\lambda'^2/\lambda^2 + 6 \lambda'^2 +1)x^2 
- \sqrt{2}(\lambda'^2/\lambda) x^3 .
\end{eqnarray}
The integration over $y'$ yields
\begin{eqnarray} \label{JA1}
J_{\sigma \sigma'}^A = I^{\sigma \sigma'} (1, 1) 
\int_0^{\frac{1}{\sqrt{2}}(1 + \lambda/\lambda')}
dx \frac{\theta(b^2(x) - 4a(x) c(x))}
{\sqrt{b^2(x) - 4a(x) c(x)}} ,   \nonumber \\
\end{eqnarray}
which can be rewritten as
\begin{eqnarray}   \label{JA3}
J_{\sigma \sigma'}^A = I^{\sigma \sigma'} (1, 1) 
\int_{-\frac{1}{\sqrt{2}}(1 + \lambda/\lambda')}
^0 dx \frac{\theta(\alpha x^2 + \beta x + \gamma)}
{\sqrt{\alpha x^2 + \beta x +\gamma}} ,
\end{eqnarray}
where $\alpha$, $\beta$, and $\gamma$, in the limit
of $\lambda \to 1, \lambda' \to 1$, can be shown as
\begin{eqnarray} \label{alpha}
\alpha= 16 ,
\end{eqnarray}
\begin{eqnarray} \label{beta}
\beta = -4 \sqrt{2}[ && \lambda \lambda' + \lambda'^2 + \lambda
+\lambda' + \lambda^2 + 3 \lambda^2 \lambda'^2  \nonumber \\
&&- 4 \lambda \lambda'^2
- 4 \lambda^2 \lambda' ] (\lambda - 1) (\lambda' -1) ,
\end{eqnarray}
\begin{eqnarray}  \label{gamma}
\gamma =-8 (\lambda -1)^2(\lambda' -1)^2 .
\end{eqnarray}
The leading order contribution to $J^A_{\sigma \sigma'}$ in fact arises
from the limiting region of $x \to 0$ in the integral of Eq. (\ref{JA3}). 
Therefore the
higher order terms of $O(x^3)$ and $O(x^4)$ have been ignored in the
$\theta$-function and the square root denominator in Eq. (\ref{JA3}).
The term $ \beta x$ can be further neglected since it also is higher
order smaller according to Eq. (\ref{beta}).
Therefore, one has
\begin{eqnarray}
J_{\sigma \sigma'}^A = && I^{\sigma \sigma'}(1, 1) 
\int_{-\frac{1}{\sqrt{2}}(1 + \lambda/\lambda')}
^0 dx    \nonumber \\ 
&& \frac{\theta(x^2 - \frac{1}{2}(\lambda -1)^2(\lambda' -1)^2)}
{4 \sqrt{x^2 - \frac{1}{2}(\lambda -1)^2(\lambda' -1)^2}} ,
\end{eqnarray}
or
\begin{eqnarray} \label{JA5}
J_{\sigma \sigma'}^A =&& \frac{1}{4} I^{\sigma \sigma'}(1, 1) 
\int_{-\frac{1}{\sqrt{2}}(1 + \lambda/\lambda')}
^{-\frac{1}{\sqrt{2}}(\lambda - 1)(\lambda' -1)} dx \nonumber \\
&& \frac{1}{\sqrt{x^2 - \frac{1}{2}(\lambda -1)^2(\lambda' -1)^2}} .
\end{eqnarray} 
Equation (\ref{JA5}) can be evaluated as
\begin{eqnarray}  \label{JAfina}
J_{\sigma \sigma'}^A = -\frac{1}{4} I^{\sigma \sigma'}(1, 1)
\ln[(\lambda - 1)(\lambda' -1)] .
\end{eqnarray}
Substituting the preceding result into Eq. (\ref{G-J}) and
carrying out the remaining integrations over $z$ and $z'$, 
one finally has
\begin{eqnarray} \label{GA}
G_{\sigma \sigma'}^A = - \frac{1}{4} I^{\sigma \sigma'}(1, 1)
(p^2 - k_F^2)^2 \ln[(p^2 - k_F^2)/2k_F^2] .
\end{eqnarray}
In view of the fact that 
$G_{\sigma \sigma'}^A$ in the preceding equation
is totally similar to
$G_{\sigma \sigma'}^{B,D}$ in Eqs. (\ref{GB}), and (\ref{GD}),
it is curious
that there seems no simpler way to derive it.

Substituting the results for $G_{\sigma \sigma'}^A$, $G_{\sigma \sigma'}^B$,
and $G_{\sigma \sigma'}^D$ in Eqs. (\ref{GA}), (\ref{GB}),
and (\ref{GD}) into Eq. (\ref{tau-G}) one finally obtains
\begin{eqnarray}
\frac{1}{\tau_e} && = \frac{\pi m}{2k_F^2}
\biggl ( \frac{1}{(2 \pi)^2} \biggr )^2 
(p^2 - k_F^2)^2 \ln[(p^2 - k_F^2)/2k_F^2]  \nonumber \\
&& \sum_{\sigma'}
[I^{\sigma \sigma'}(1, 1)  
+ I^{\sigma \sigma'}(1, -1)
+ I^{\sigma \sigma'}(-1, 1) ] .
\end{eqnarray}
The contributions from the ``direct" and ``exchange" processes can 
be separately written as
\begin{eqnarray}
\frac{1}{\tau^{(D)}}= && - \frac{m^2 \xi_p^2}{16 \pi^3 E_F}
\ln \frac{\xi_p}{2E_F}     \nonumber \\
&& \sum_{\sigma'} [2 (W_{\sigma \sigma'}(0))^2
+(W_{\sigma \sigma'}(2k_F))^2 ] ,
\end{eqnarray}
and
\begin{eqnarray}
\frac{1}{\tau^{(ex)}}=&& \frac{m^2 \xi_p^2}{16 \pi^3 E_F}
\ln \frac{\xi_p}{2E_F}   \nonumber \\
&& [(W_{\sigma \sigma}(0))^2 
+ 2W_{\sigma \sigma}(0)W_{\sigma \sigma}(2k_F)] .
\end{eqnarray}
We empahsize that the above results are accurate only
to the leading order of $O(\xi_p^2 \ln \xi_p)$.
In the case that the spin dependence of the effective interaction
can be neglected, one has
exactly the results shown in Eqs. (52) and (59) in I, respectively.
The sum of $1/\tau^{(D)}$ and $1/\tau^{(ex)}$ yields $1/\tau_e$
as given in Eq. (2) in the introduction.

\section{Discussion and summary}

\begin{figure}
\unitlength1cm
\begin{picture}(5.0,6.0)
\put(-5.0,-4.0){\makebox(7.0,8.0){
\includegraphics{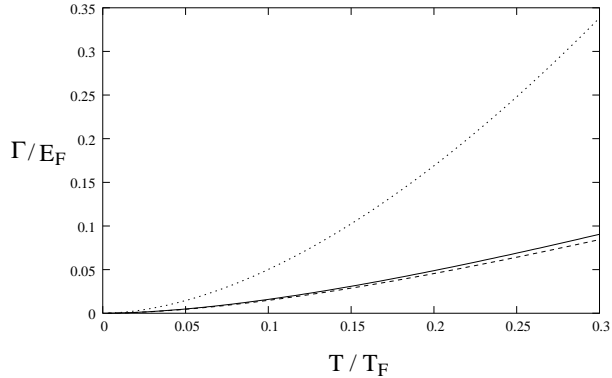}
}}
\end{picture}
\caption{Electron relaxation rate $\Gamma$ 
in 2D.\cite{murphy,QV,jungwirth,zheng,reizer} 
Dotted line: experimental
data from Ref. \cite{murphy}; dashed line: calculated one from Eq. (3)
with RPA to the effective interaction $W(q)$;\cite{QV}
solid line: calculated one from Eq. (3) with the effects of
the local field and the finite well width on $W(q)$ included.}
\label{figur2}
\end{figure}
The results in Eqs. (\ref{tau3D}) and (\ref{tau3ex}) 
have been obtained with
the approximation of the Thomas-Fermi screened Coulomb 
potential of Eq. (\ref{TF})
to the effective electron interaction. One can always resort to the
more general expressions of Eqs. (\ref{3DDgen}) and (\ref{3Dexgen}) 
if necessary. On the other side, in the 2D case, with which this
paper is mainly concerned,
no approximation has been made in the effective 
electron interaction $W(q)$ except that it is assumed to be static. The local
field effects and the finite well width effects on 
the effective interaction can be readily
taken into account. We now estimate their contribution to
the inverse quasiparticle lifetime by using the form factor of
$F(q)=\frac{2}{qb} [1+\frac{1}{qb}(e^{-qb} - 1) ]$ with $b$ being the
well width,\cite{zheng,macdonald} and the local field factor evaluated
in Ref. \cite{g-factor}. These effects have been investigated earlier
in Ref. \cite{zheng}. The clarification of the 
theoretical disagreement now enables us to definitely elucidate their
contribution. Both of them are shown to 
yield in effect quite small corrections to the results calculated with
random phase approximation (RPA) to $W(q)$. 
This conclusion should hold in more general sense regardless of particular
choice of the form factor and the local field factor, since both of them
mainly effect the short-range behavior of the effective 
interaction while the inverse lifetime 
of a low energy quasiparticle is mainly determined 
by the long-range behavior of the
effective interaction. The comparison with the experimental values of the
electron relaxation 
rate $\Gamma$ \cite{murphy,QV,jungwirth,zheng,reizer} 
from Ref. \cite{murphy} is illustrated
in Fig. 2. The result calculated with the RPA to the effective
interaction is also plotted in Fig. 2 for comparision.\cite{QV}

It seems that other factors must be taken into account 
in order to explain the difference
between theory and experiment. One of the assumptions made in all
previous work is that the couplings between electrons in different
wells are weak and can be ignored. This assumption might require
further justication
for a barrier width being about $250 \AA$, while the 
width of each well being about
$200 \AA$. Furthermore, higher order terms in electron interaction, 
usually not important
at a density of $r_s \sim 1$ where $r_s$ is the Wigner-Seitz radius, 
might not be simply ignored 
in this case, for they have been
shown to contribute nontrivially higher-order 
logarithmic factors.\cite{menashe}
Other factors which might also play a role have been metioned in Ref. \cite{QV}.  
Evidently, further theoretical effort is needed in order to fully understand the
discrepancy between theory and experiment.
 
In conclusion, we have calculated, in a rather different manner, the
inelastic lifetime of an electron quasiparticle in an
electron liquid. The results confirm those in Eq. (\ref{tau3Dfina}), 
and Eqs. (\ref{taueeT=0})
and (\ref{taueeT>0}) obtained in our previous work, and consequently
finally resolve the theoretical discrepancies in the literature.

\section{Acknowledgments}
The author is grateful to Prof. G. Vignale
for advice and discussions.
This work was supported by the Chinese National Science Foundation
under Grant No. 10474001.

\end{document}